\documentclass[english,aps,twocolumn,showpacs,prl,superscriptaddress,floatfix,longbibliography]{revtex4-1}

\usepackage{bm}
\usepackage{hyperref}
\usepackage{amsmath}
\usepackage{amssymb}
\usepackage{color}
\usepackage{xspace} 
\usepackage{graphicx}
\usepackage{color}
\usepackage{epsfig}
\usepackage{epstopdf}

\begin{document}

\title{Origins of the unidirectional spin Hall magnetoresistance in metallic bilayers}

\author{Can Onur Avci}
\affiliation{Department of Materials Science and Engineering, Massachusetts Institute of Technology,
Cambridge, Massachusetts 02139, USA}
\author{Johannes Mendil}
\affiliation{Department of Materials, ETH Z{\"u}rich, 8093 Z{\"u}rich, Switzerland}
\author{Geoffrey S. D. Beach}
\affiliation{Department of Materials Science and Engineering, Massachusetts Institute of Technology,
Cambridge, Massachusetts 02139, USA}
\author{Pietro Gambardella}
\affiliation{Department of Materials, ETH Z{\"u}rich, 8093 Z{\"u}rich, Switzerland}


\begin{abstract}
Recent studies evidence the emergence of asymmetric electron transport in layered conductors owing to the interplay between  electrical conductivity, magnetization, and the spin Hall or Rashba-Edelstein effects. Here, we investigate the unidirectional magnetoresistance (UMR) caused by the current-induced spin accumulation in Co/Pt and CoCr/Pt bilayers. We identify three competing mechanisms underpinning the resistance asymmetry, namely interface and bulk spin-dependent electron scattering and electron-magnon scattering. Our measurements provide a consistent description of the current, magnetic field, and temperature dependence of the UMR and show that both positive and negative UMR can be obtained by tuning the interface and bulk spin-dependent scattering terms relative to the magnon population.
\end{abstract}
\maketitle

The interconversion of charge and spin currents is a central theme in spintronics. The spin accumulation generated by the spin Hall effect (SHE) and/or the Rashba-Edelstein effect (REE) enables efficient current-induced magnetization switching, domain wall manipulation, and ferromagnetic resonance \cite{Manchon2018}.
Moreover, the coupling between the spin and orbital moments of the charge carriers, as exemplified by the SHE and REE, is responsible for novel magnetoresistive phenomena, such as the spin Hall magnetoresistance (SMR) \cite{Nakayama2013,Althammer2013,Vlietstra2013,Chen2013b,Miao2014,Isasa2014,Avci2015a,Avci2015b,Kim2016},
the Hanle magnetoresistance \cite{Dyakonov2007,Velez2016}, and the Rashba-Edelstein magnetoresistance (EMR) \cite{Kobs2011,Grigoryan2014,Zhang2014Zhang,Zhang2015e,Hupfauer2015,Nakayama2016,Nakayama2017,Kim2017f}.
These phenomena are transforming our understanding of electric transport, leading to novel possibilities to sense the magnetization in devices. The archetypal SMR, for example, arises from the conversion of a charge current density $\mathbf{j} \parallel \mathbf{x}$ flowing in the plane of a ferromagnet/normal metal (FM/NM) bilayer into a spin current diffusing along $\mathbf{z}$ into the FM, with spin polarization $\bm{\sigma} \parallel \mathbf{y}$. For parallel or antiparallel orientation of $\bm{\sigma}$ and magnetization $\mathbf{m}$, part of the spin current is reflected at the FM/NM interface and back-converted into a charge current by the inverse SHE. The additional electric current arising from the combination of the direct and inverse SHE leads to a reduction of the resistance proportional to $m_y^2$ \cite{Chen2013b}. The EMR has the same symmetry as the SMR, but is attributed to spin mixing due to the interfacial REE \cite{Grigoryan2014,Zhang2014Zhang,Zhang2015e}.


\begin{figure}[b]
	\includegraphics[width=\columnwidth]{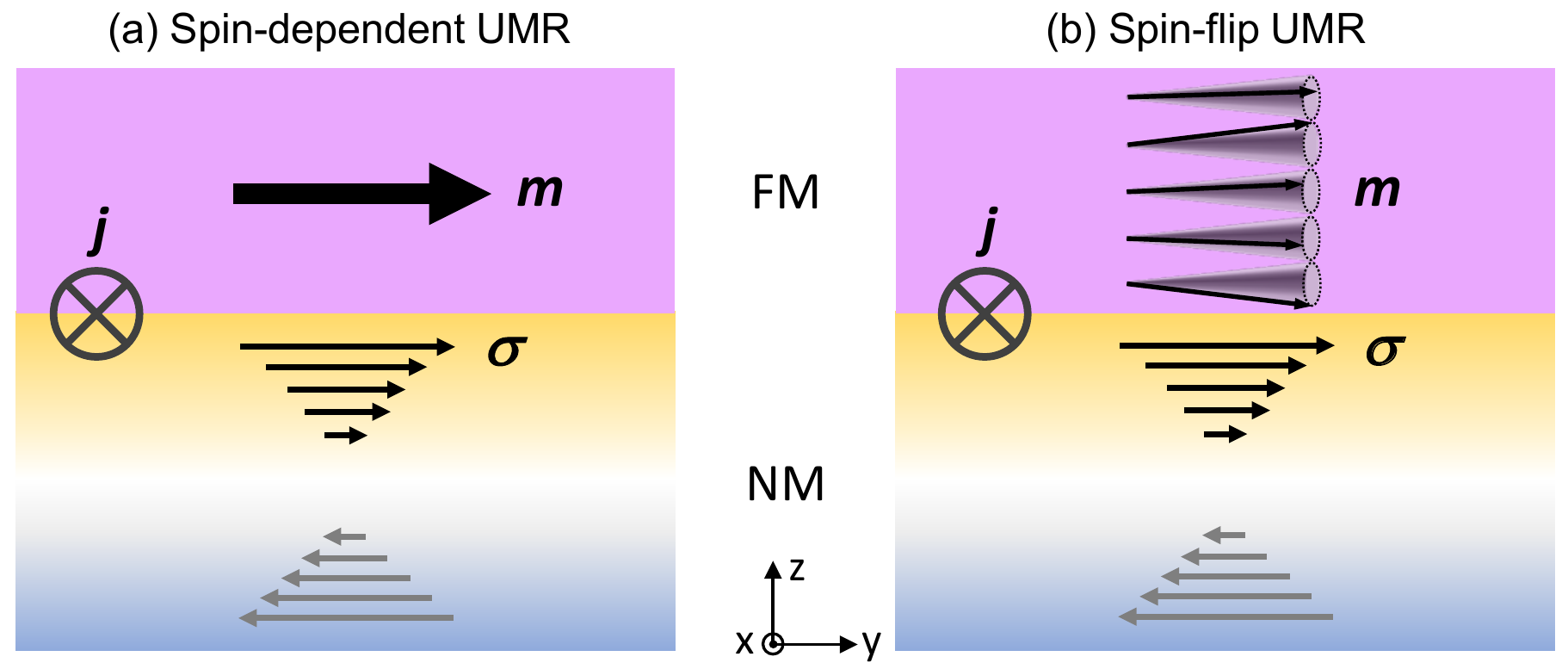}
	\caption{(a) Illustration of the spin-dependent UMR and (b) spin-flip SMR.}
	\label{fig1}
\end{figure}

\begin{figure}[t]
	\includegraphics[width=\columnwidth]{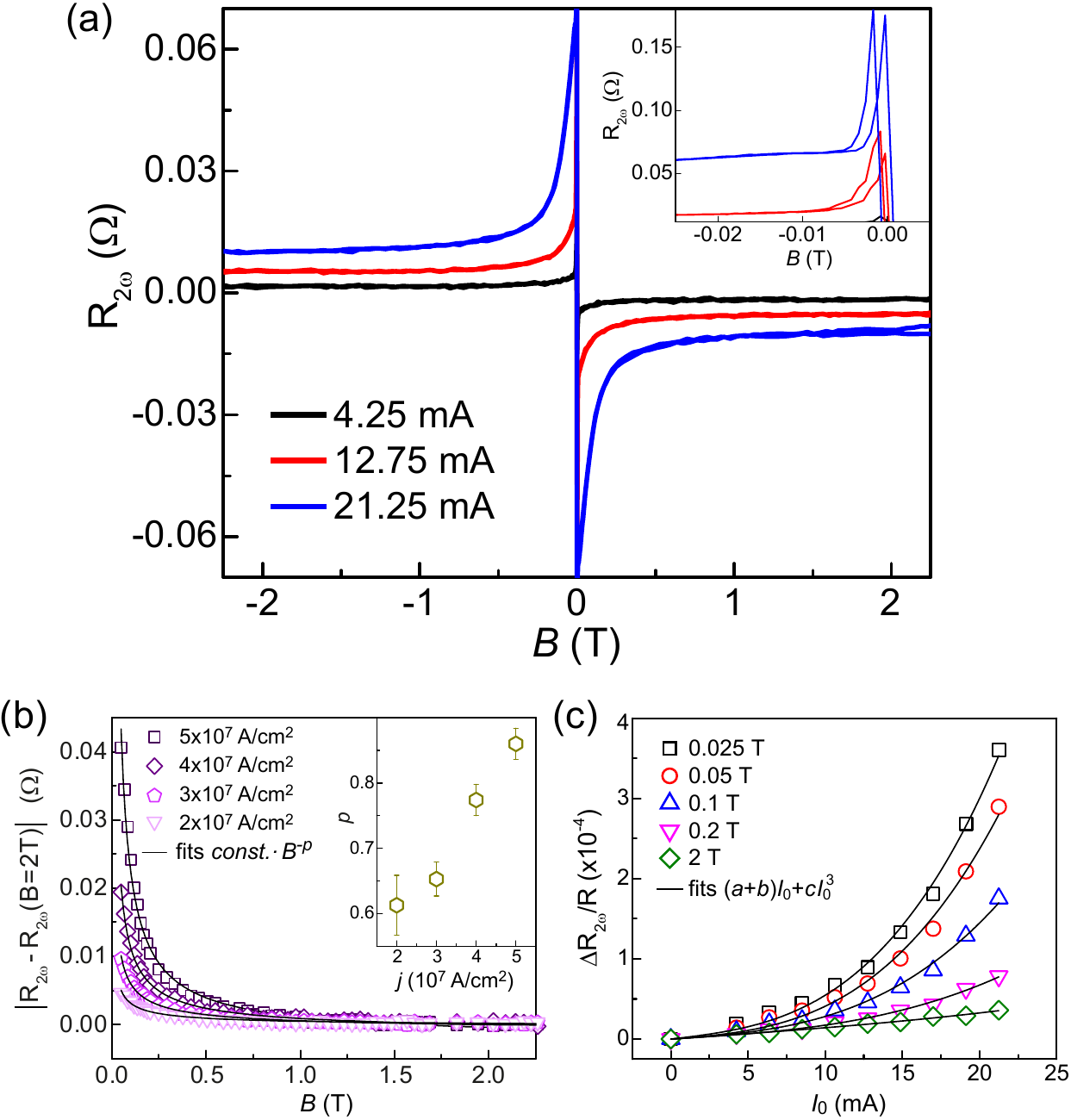}
	\caption{(a) Unidirectional magnetoresistance $R_{2\omega}$ as a function of magnetic field and current. Inset: detail of the low field region. (b) Fits of the field-dependent contribution to the UMR, $|R_{2\omega}-R_{2\omega}(B=2\,\mathrm{T})|\propto B^{-p}$, at different currents. Inset: dependence of the exponent $p$ on current. (c) Fits of the current dependence of the UMR at different fields.}
\label{fig2}
\end{figure}

Recent studies have shown that an additional current-dependent unidirectional magnetoresistance (UMR) emerges in FM/NM systems due to either the SHE or REE \cite{Avci2015a,Avci2015b,Olejnik2015,Langenfeld2016,Ono2017,Yin2017,Avci2017b,Yasuda2016,Yasuda2017}.
Unlike the most common magnetoresistive effects, including the anisotropic magnetoresistance (AMR), SMR, and EMR, the UMR is a nonlinear effect that violates Onsager reciprocity, being odd under either magnetization or current reversal. Interestingly, the UMR is a general property of FM/NM, FM/semiconductor, and FM/topological insulator bilayers~\cite{Avci2015a,Avci2015b,Langenfeld2016,Ono2017,Yin2017,Avci2017b,Olejnik2015,Yasuda2016,Yasuda2017,Lv2018}.
Because of its relationship to spin-charge conversion and electron scattering effects, the UMR provides fundamental insight into the transport properties of spin-orbit coupled systems, including bulk crystals \cite{Ideue2017}. Moreover, owing to its unidirectional properties, the UMR can be used to electrically detect the sign of the magnetization in bilayer and multilayer samples using a simple two-terminal geometry \cite{Avci2015a,Olejnik2015,Avci2017b}. However, despite this intense interest, the microscopic origins of the UMR are still under debate.

Different mechanisms can give rise to UMR in FM/NM systems, even when considering a single source of spin accumulation such as the SHE. A first mechanism, sketched in Fig.~\ref{fig1}(a), is the modulation of the interface resistance between the FM and NM due to the SHE-induced spin polarization, which changes the reflection and transmission coefficients of the electrons depending on the orientation of $\bm{\sigma}$ relative to $\mathbf{m}$ \cite{Avci2015a}. A second mechanisms relies on the bulk spin-dependent conductivity of the FM, which again enhances or decreases the resistance of the FM/NM bilayer for parallel and antiparallel alignment of $\bm{\sigma}$ and $\mathbf{m}$ \cite{Zhang2017a}. Both such mechanisms find a strong analogy with the current-in-plane giant magnetoresistance (GMR) in FM/NM/FM trilayers \cite{Baibich1988,Binasch1989,Camley1989,Hood1992}, where the role of one FM polarizer is replaced by the SHE in the NM, and differ from one another in the crucial role played by spin-dependent scattering occurring at the interface or in the bulk of the FM. A third mechanism, sketched in Fig.~\ref{fig1}(b), invokes the creation or annihilation of magnons resulting from the absorption of the SHE-induced spin current in the FM \cite{Demidov2011b}. In this case, the spin flips caused by electron-magnon scattering result in an increase or decrease, respectively, of the longitudinal resistance of the FM \cite{Langenfeld2016,Ono2017,Yasuda2016}. This last mechanism is related to the so-called spin-disorder resistivity of single FM conductors \cite{Kasuya1956,Goodings1963}, which emerges also in the temperature and field-dependent measurements of thin films \cite{Raquet2002,Mihai2008,Rowan2014,Cheng2017}. Note that these mechanisms differ from the non-local MR recently reported in magnetic insulators/NM due to pure magnon currents \cite{Cornelissen2015,Goennenwein2015}.
In the following, we refer to the first and second mechanism as the interface and bulk spin-dependent (SD) UMR, respectively, and to the third as spin-flip (SF) UMR.

In this work, we investigate the origin of the UMR in FM/NM metal layers as well as its current, magnetic field, and temperature dependence. We find that the three mechanisms described above coexist in Co/Pt bilayers and that the SD-UMR and SF-UMR can be separated according to their different field and current dependence. Measurements of Co$_{80}$Cr$_{20}$/Pt further show that the interface and bulk spin-dependent scattering can be independently tuned to determine the sign and magnitude of the SD-UMR, similar to the direct and inverse GMR effect, whereas the SF-UMR depends on the temperature and magnon stiffness of the FM layer. Our results provide a unified picture of the microscopic processes leading to nonreciprocal electric transport in FM/NM conductors as well as practical insight on how to design heterostructures with tunable UMR.

We studied multilayer samples consisting of Ta(2.5)/Co(2.5)/Pt(6)/Ta(2)/substrate and Ta(2.5)/Co$_{80}$Cr$_{20}$(1.6-5)/Pt(4)/Ta(2)/substrate grown on thermally oxidized Si wafers by magnetron sputtering (numbers in parentheses are thicknesses in nm). The top Ta layer is naturally oxidized and nonconducting, and we assume that current shunting by the bottom Ta seed layer is negligible due to its high resistivity. All samples have in-plane magnetization. The blanket layers were patterned by optical lithography into Hall bars with lateral width $w=5\, , 10$~$\mu$m and length $l\approx 4w$. The MR measurements were performed by applying an ac current $I=I_0 \sin(\omega t)$ of frequency $\omega/2\pi=10$~Hz and recording the first and second harmonic of the longitudinal voltage $V=V_{\omega} + V_{2\omega}= IR_{\omega} + IR_{2\omega}$ as a function of $I$ and external magnetic field $B$ \cite{Avci2015a}. Here, the first harmonic $R_{\omega}$ represents the usual current-independent resistance of the bilayer, which includes contributions from the AMR, SMR, and EMR.
The second harmonic $R_{2\omega}(I)$ includes the different current-dependent contributions to the resistance, namely the UMR, the changes of the MR due to the oscillation of $\mathbf{m}$ induced by the spin-orbit torques \cite{Garello2013}, and the magnetothermal voltage induced by temperature gradients \cite{Avci2014b}. These contributions can be distinguished by their different symmetry and field dependence \cite{Avci2015a,SI}. In Co(2.5)/Pt(6), the magnetothermal voltage is less than 5 \% of the total signal and the spin-orbit torque-induced oscillations of the MR are null for $\mathbf{m} \parallel \mathbf{y}$ \cite{SI}. In these conditions, $R_{2\omega}$ represents the change of resistance for positive and negative applied current, namely the UMR.

Figure~\ref{fig2} (a) shows $R_{2\omega}$ as a function of applied field $\mathbf{B} \parallel \mathbf{y}$ and $I$. The data evidence two distinct regimes, corresponding to low and high values of $B$. Above 1~T, $R_{2\omega}$ is dominated by a constant term that is independent of $B$ and proportional to $I$. This term is the SD-UMR previously reported by us and other groups \cite{Avci2015a,Avci2015b,Olejnik2015,Yin2017,Zhang2017a}. Below 1~T, on the other hand, $|R_{2\omega}|$ increases sharply following a power law $|R_{2\omega}(B,I)-R_{2\omega}(2~T,I)| \propto B^{-p}$, with $p$ varying monotonically from 0.6 to 0.9 as a function of increasing current [Fig.~\ref{fig2}(b)]. The increase of $R_{2\omega}$ is even more remarkable near zero field [inset of Fig.~\ref{fig2}(a)]. However, as the magnetization is not uniform and is hysteretic in this limit, our analysis focuses on fields $|B|>0.02$~T. These data provide a first indication that different mechanisms simultaneously contribute to the UMR. In order to gain further insight into such mechanisms, we fit the relative resistance change $\Delta R_{2\omega}/R = [R_{2\omega}(\mathbf{m}\parallel +\mathbf{y})-R_{2\omega}(\mathbf{m}\parallel -\mathbf{y})]/R$ with a polynomial function of the current [Fig.~\ref{fig2}(c)]. We find excellent agreement for an expression of the type $[a +b(B)]I + c(B)I^3$, where $a$ is a constant independent of $B$, and $b$ and $c$ are two coefficients that scale inversely with $B$. This expression again supports the presence of two distinct scattering processes, one scaling with $aI$ and the other with $b(B)I+c(B)I^3$. Whereas the first term is consistent with the SD-UMR, the field dependence of the remaining terms indicate that the second process is related to the magnon population in the Co layer. It is well known that an applied field strongly reduces the magnon density in thin films, leading to a decrease of the resistance due to the reduction of thermal spin disorder \cite{Raquet2002}. Such an effect is clearly present also in our samples, and influences both $R_{\omega}$ \cite{SI} and $R_{2\omega}$. We therefore attribute the decrease of the UMR in Fig.~\ref{fig2} to the field counteracting the excitation of magnons by the spin current.

Further support for an electron-magnon scattering mechanism comes from the nonlinear current dependence shown in Fig.~\ref{fig2}(c). Strong nonlinearities in the magnon population have been observed by Brillouin light scattering as the current intensity approaches the damping compensation threshold in FM/NM bilayers \cite{Demidov2011b}. Together with Joule heating, such nonlinear effects determine the nonequilibrium density of magnons in the FM \cite{Demidov2011b,Meyer2017}, which ultimately affects $R_{2\omega}$ due to spin-flip processes. In Co/Pt, our fits of the current dependence suggest that the spin current ($\propto I$) modulates a thermalized magnon population $\propto (T+\Delta T)\propto (b(B)+c(B)I^2)$, where $T$ is the ambient temperature and $\Delta T \propto I^2$ is the temperature increase due to Joule heating \cite{SI}. We thus conclude that the UMR is given by the concurrence of spin-dependent and spin-flip scattering processes that have very different field and current dependencies. Whereas the SD-UMR dominates at high field, the SF-UMR produces the strongest MR asymmetry at low field and high current. These results reconcile the interpretation of the UMR in terms of spin-dependent conductivity \cite{Avci2015a,Zhang2017a} and magnon excitations \cite{Langenfeld2016,Ono2017,Yasuda2016}.

\begin{figure}[t]
	\includegraphics[width=\columnwidth]{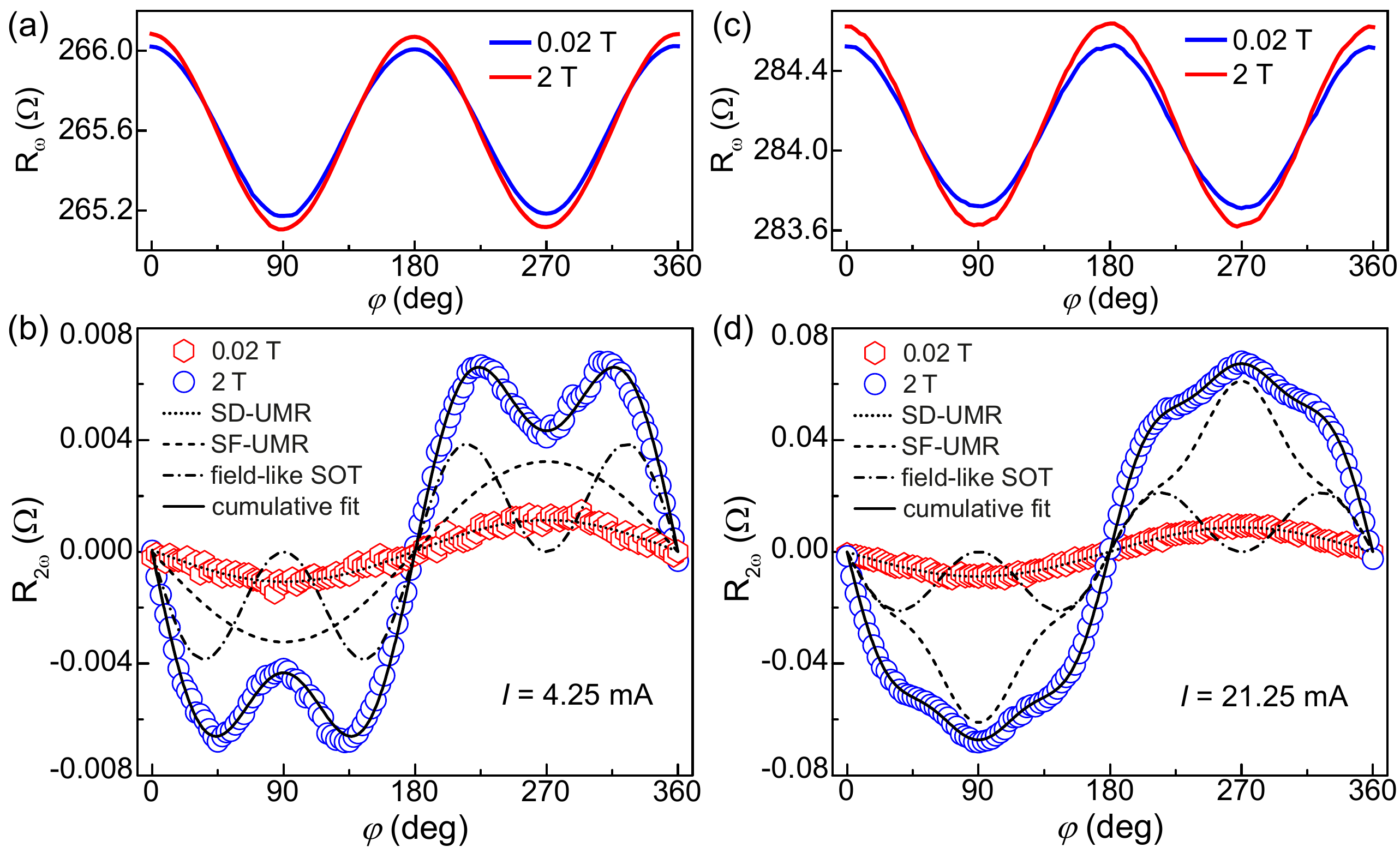}
	\caption{Angular dependence of (a) $R_{\omega}$ and (b) $R_{2\omega}$ at low current ($j=1 \times 10^7$A/cm$^{2}$) and (c,d) high current ($j=5\times 10^7$A/cm$^{2}$). The dotted, dot-dashed, and dashed lines are fits of the SD-UMR, spin-orbit torque, and SF-UMR contributions to $R_{2\omega}$, respectively. See Ref.~\onlinecite{SI} for a description of the fit procedure.}
	\label{fig3}
\end{figure}

\begin{figure*}[t]
	\includegraphics[width=0.8\textwidth]{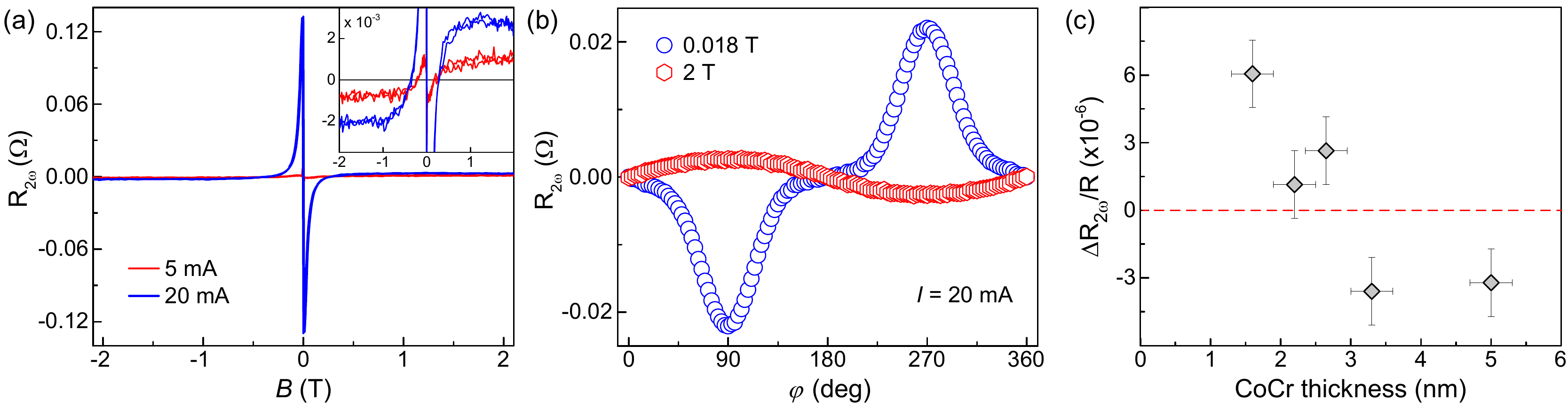}
	\caption{(a) UMR of Co$_{80}$Cr$_{20}$(3.3)/Pt(4) as a function of magnetic field $B \parallel \mathbf{y}$. The curves show $R_{2\omega}(B,I)$ measured at $I = 5$~mA ($j = 0.7\times 10^7$ A/cm$^2$) and 20~mA ($2.7\times 10^7$ A/cm$^2$). Inset: detail of the sign reversal of $R_{2\omega}(B,I)$ as the negative SD-UMR prevails over the SF-UMR at high field.(b) Angular dependence of $R_{2\omega}(B,I)$ at low and high field, representing the SF- and SD-UMR, respectively. (c) SD-UMR as a function of Co$_{80}$Cr$_{20}$ thickness measured at $B=2$~T and $j = 1\times 10^7$ A/cm$^2$. The dashed line indicates the compensation of the bulk and interface spin-dependent scattering.}
	\label{fig4}
\end{figure*}

Temperature- and angular-dependent measurements of $R_{2\omega}$ offer further insight into the different properties of the SD-UMR and SF-UMR. In the Supplementary Material \cite{SI}, we show that the SF-UMR decreases almost 10-fold from 300~K to 4~K, whereas the SD-UMR decreases only 2-fold, highlighting the prominent role played by magnons in the first effect. Figure~\ref{fig3} shows the angular-dependence of $R_{\omega}$ and $R_{2\omega}$ measured at constant $B$ and $I$ while rotating the field in the $xy$ plane by an angle $\varphi$. We find that $R_{\omega}(\varphi)$ is proportional to $\sin^2 \varphi \propto m_y^2$ and not significantly affected by either $B$ or $I$, as expected for the SMR and AMR of a magnetically saturated layer [Fig.~\ref{fig3}(a) and (c)]. On the other hand, $R_{2\omega}(\varphi)$ varies strongly between low and high field and also between low and high current. At high field both the spin-orbit torque and SF-UMR signals are small, and we observe the typical $\sin \varphi$ behavior expected of the SD-UMR [red symbols and dotted line in Fig.~\ref{fig3}(b) and (d)]. At low current and low field [blue symbols in Fig.~\ref{fig3}(b)] we observe four peaks at $\varphi=45^{\circ}, 135^{\circ}, 225^{\circ}, 315^{\circ}$, which are characteristic of the field-like spin-orbit torque and Oersted field contribution to $R_{2\omega}(\varphi)$ (dot-dashed line) superimposed on the UMR \cite{Avci2015a,Avci2014b}. At high current and low field [blue symbols in Fig.~\ref{fig3}(d)], the SF-UMR signal is strongest, which results in two peaks at $\varphi=90^{\circ}$ and $270^{\circ}$ (dashed line). Interestingly, the high current SF-UMR is not simply proportional to $m_y \propto \sin \varphi$ as expected based on the product $\mathbf{m} \cdot \mathbf{\sigma}$, but strongly peaked around $\varphi=90^{\circ}$ and $270^{\circ}$. Such a peaked angular dependence, which is even more evident for Co$_{80}$Cr$_{20}$/Pt [see Fig.~\ref{fig4}], suggests that the magnon excitation probability becomes anisotropic as the current approaches the damping compensation threshold in the FM. We have presently no model for this effect, but note that such an anisotropy cannot be excluded on theoretical grounds \cite{Rezende2005} and that a peaked angular dependence has been reported also in the spin pumping signal of Pt/YIG bilayers \cite{Collet2016}.

%

Finally, we show that the SD-UMR consists of two separate contributions arising from bulk and interface spin-dependent scattering.
In analogy with the GMR, we define a positive UMR when $\bm{\sigma}$ is parallel to the majority spins in the FM, resulting in a low resistance state. This situation is the most common and occurs, e.g., in Co/Pt, Co/Ta, and Co/W bilayers, for which we have confirmed the sign of $\bm{\sigma}$ by spin-orbit torque measurements. However, it is well known that a negative (inverse) GMR can be realized in magnetic multilayers in which the spin asymmetry coefficients for bulk ($\beta$) and interface ($\gamma$) scattering have opposite sign, such as (FeCr, FeV, CoCr)/Cu/Co \cite{George1994,Vouille1999}. Because spin-dependent scattering underpins both phenomena \cite{Zhang2017a}, we expect that the magnitude and sign of the SD-UMR can be tuned in a similar way as the GMR. Further, by comparing the SD-UMR in systems with opposite sign of $\beta$ and $\gamma$, it should be possible to separately determine the bulk and interface contributions to the SD-UMR.

To test these hypotheses, we chose Co$_{80}$Cr$_{20}$ as a model FM in which the conductivity of the minority electrons is larger than that of the majority electrons, i.e., $\beta < 0$ \cite{Vouille1999}, and measured the UMR of Co$_{80}$Cr$_{20}$/Pt(4) bilayers of different thickness. Figure~\ref{fig4}(a) shows the $R_{2\omega}$ of Co$_{80}$Cr$_{20}$(3.3)/Pt(4). Similar to the measurements reported in Fig.~\ref{fig2}(c), we observe that $R_{2\omega}$ is significantly enhanced at high current and low field. This enhancement arises from the SF-UMR, which has the same sign as in Co/Pt. In contrast to Co/Pt, however, $R_{2\omega}$ changes sign above 0.15~T, becoming negative in the high field regime dominated by the SD-UMR. The sign reversal is confirmed by the angular-dependent measurements of $R_{2\omega}$ performed at fields representative of the SF-UMR and SD-UMR regimes [Fig.~\ref{fig4}(b)]. We thus focus on the high field behavior of $R_{2\omega}$ in Co$_{80}$Cr$_{20}$ to investigate the sign change of the SD-UMR. Figure~\ref{fig4}(c) reports $\Delta R_{2\omega}(2 T)/R$ as a function of Co$_{80}$Cr$_{20}$ thickness after subtraction of the magnetothermal signal \cite{SI}. The relatively large error bars are due to low signal-to-noise ratio, uncertainties in the separation of the magnetothermal and UMR voltages, and thickness variations along the Co$_{80}$Cr$_{20}$ wedge. We observe that $\Delta R_{2\omega}(2 T)/R$ is positive below $\sim 3$~nm, similar to Co/Pt, and negative above. The existence of a compensation thickness with zero UMR unambiguously demonstrates that $\beta$ and $\gamma$ have opposite sign in Co$_{80}$Cr$_{20}$/Pt and that the SD-UMR of the thicker films is determined by bulk spin-dependent scattering with $\beta < 0$. Such a behavior is reminiscent of the GMR inversion in Co$_{80}$Cr$_{20}$/Cu/Co multilayers \cite{Vouille1999}, which leads us to conclude that there are two competing contributions to the SD-UMR: one due to interface scattering, which is generally positive ($\gamma >0$) and prevails in the limit of thin FM, and one due to bulk scattering, which can be either positive ($\beta >0$) or negative ($\beta <0$) and dominates in thick FM.

A corollary to these measurements is that the SF-UMR has the same sign in Co$_{80}$Cr$_{20}$/Pt as in Co/Pt, independently of thickness. This result can be easily explained by considering that the direction of the spin-orbit torques and magnetization remain the same in the two systems, such that the combination of current and magnetization required for exciting or annihilating magnons does not change. Interestingly, however, the field-induced damping of the SF-UMR does depend on the Co$_{80}$Cr$_{20}$ thickness. For this system, we find that $|R_{2\omega}(B,I)-R_{2\omega}(2~T,I)| \propto B^{-p}$, with $p$ dropping from 1.7 to 1.1 in $~2$~nm and 5~nm thick Co$_{80}$Cr$_{20}$ films, respectively \cite{SI}. Such a drop may be attributed to the increase in magnon stiffness that occurs in FM films as spin disorder progressively reduces with increasing thickness \cite{Rowan2014}. This behavior provides additional evidence that the SF-UMR and SD-UMR originate from distinct phenomena and can be separately controlled by modifying the composition and thickness of the FM layer.

In summary, we have shown that three different mechanisms determine the UMR of metal bilayers, namely the bulk and interface SD-UMR and the SF-UMR. These mechanisms can be separated by their distinct field and current dependence. Whereas the SD-UMR is independent of $B$ and proportional to $j$, the SF-UMR scales with $B^{-p}$ and is proportional to $j +j^3$. The monotonic field dependence of the SF-UMR originates from the field-induced gap in the magnon excitation spectrum, which quenches the electron-magnon scattering at high field.
The exponent of the power law $B^{-p}$ increases with current and decreases with the thickness of the FM \cite{SI}, as expected for the softening of the magnon modes with temperature due to Joule heating and stiffening of the modes with thickness, respectively. Another prominent difference between the SF-UMR and SD-UMR is that the former is always positive, whereas the latter can be either positive or negative. The positive SD-UMR of Co/Pt concords with the positive spin asymmetry coefficients for bulk and interface scattering of Co and Co/Pt, respectively, as determined by GMR \cite{Vouille1999,Nguyen2014}. Measurements of Co$_{80}$Cr$_{20}$/Pt, on the other hand, show that the SD-UMR becomes negative when Co$_{80}$Cr$_{20}$ is thicker than 3 nm. This behavior is similar to the inverse GMR effect,
which indicates that both the interface and bulk SD-UMR are present and have opposite sign.
The possibility of tuning the UMR by modifying the magnon spectrum as well as the relative weight of bulk and interface electron scattering makes this phenomenon very appealing to study electron transport in spin-orbit coupled systems as well as to measure the magnetization in two-terminal devices.

We acknowledge funding by the Swiss National Science Foundation (grant No.\ 200020-172775) and by C-SPIN, one of the six SRC STARnet Centers, sponsored by MARCO and DARPA.

%

\end{document}


\includepdf[pages=-]{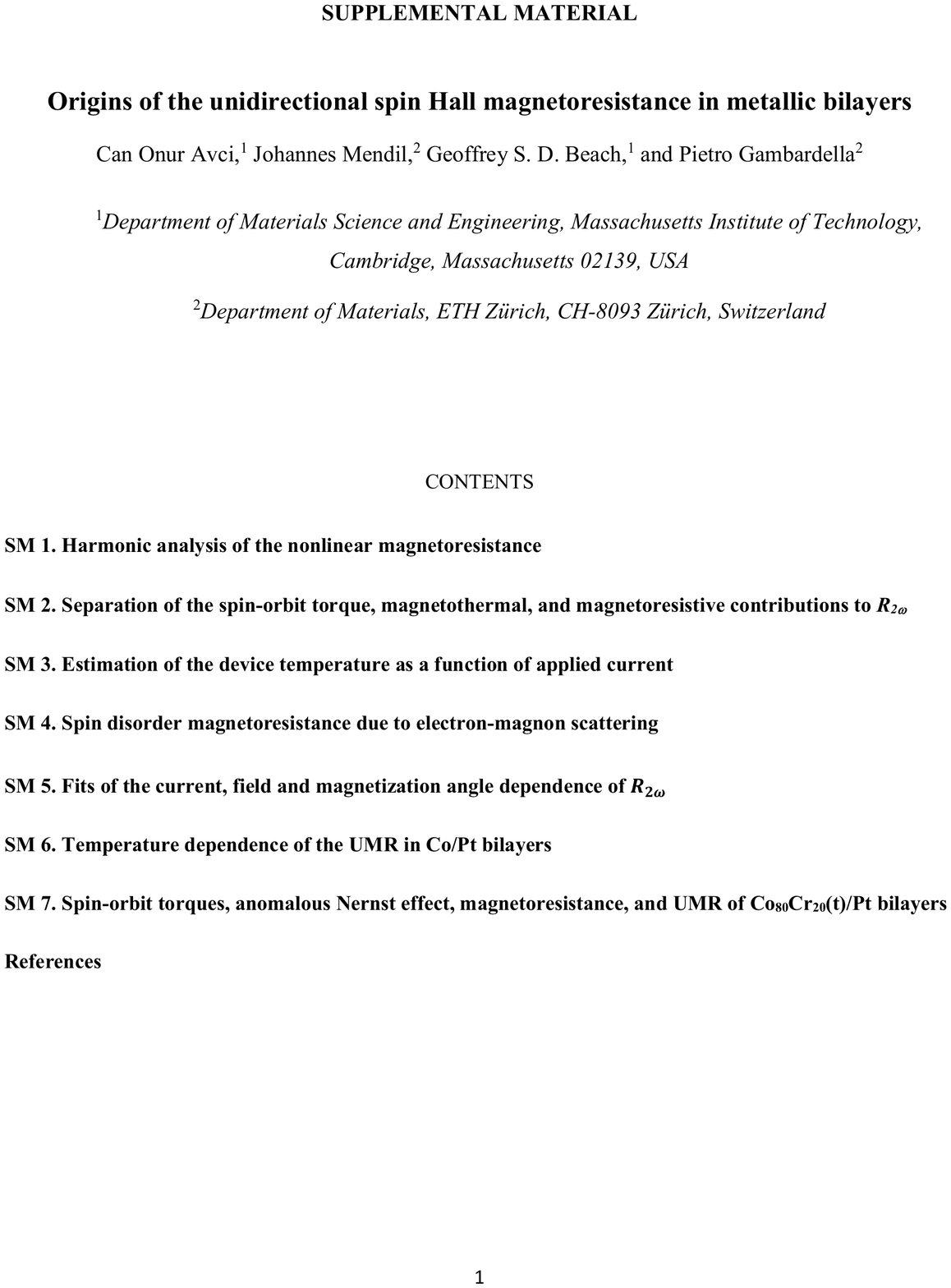}
\includepdf[pages=-]{Avci_SM_Part1.pdf}
\includepdf[pages=-]{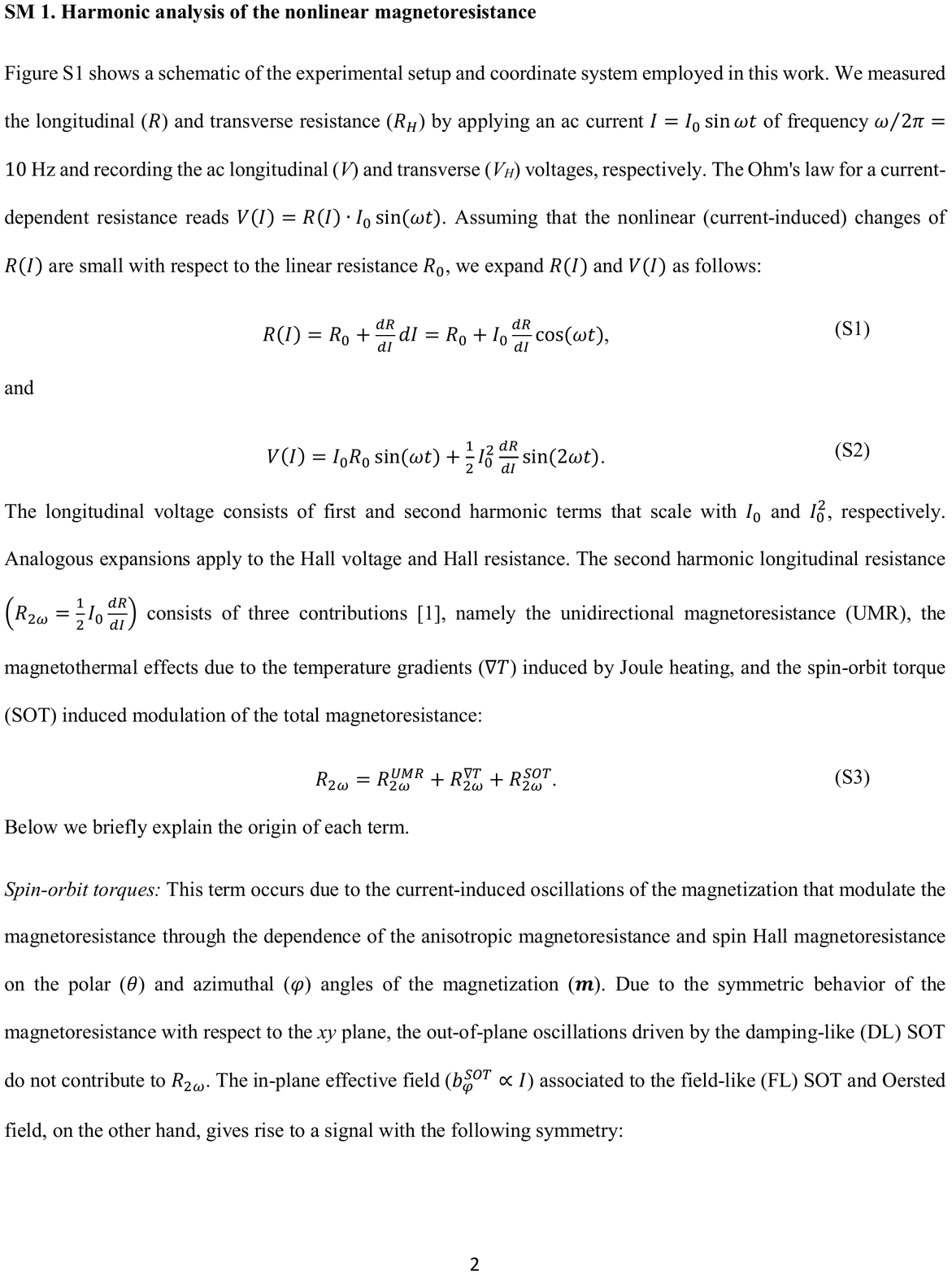}
\includepdf[pages=-]{Avci_SM_Part2.pdf}
\includepdf[pages=-]{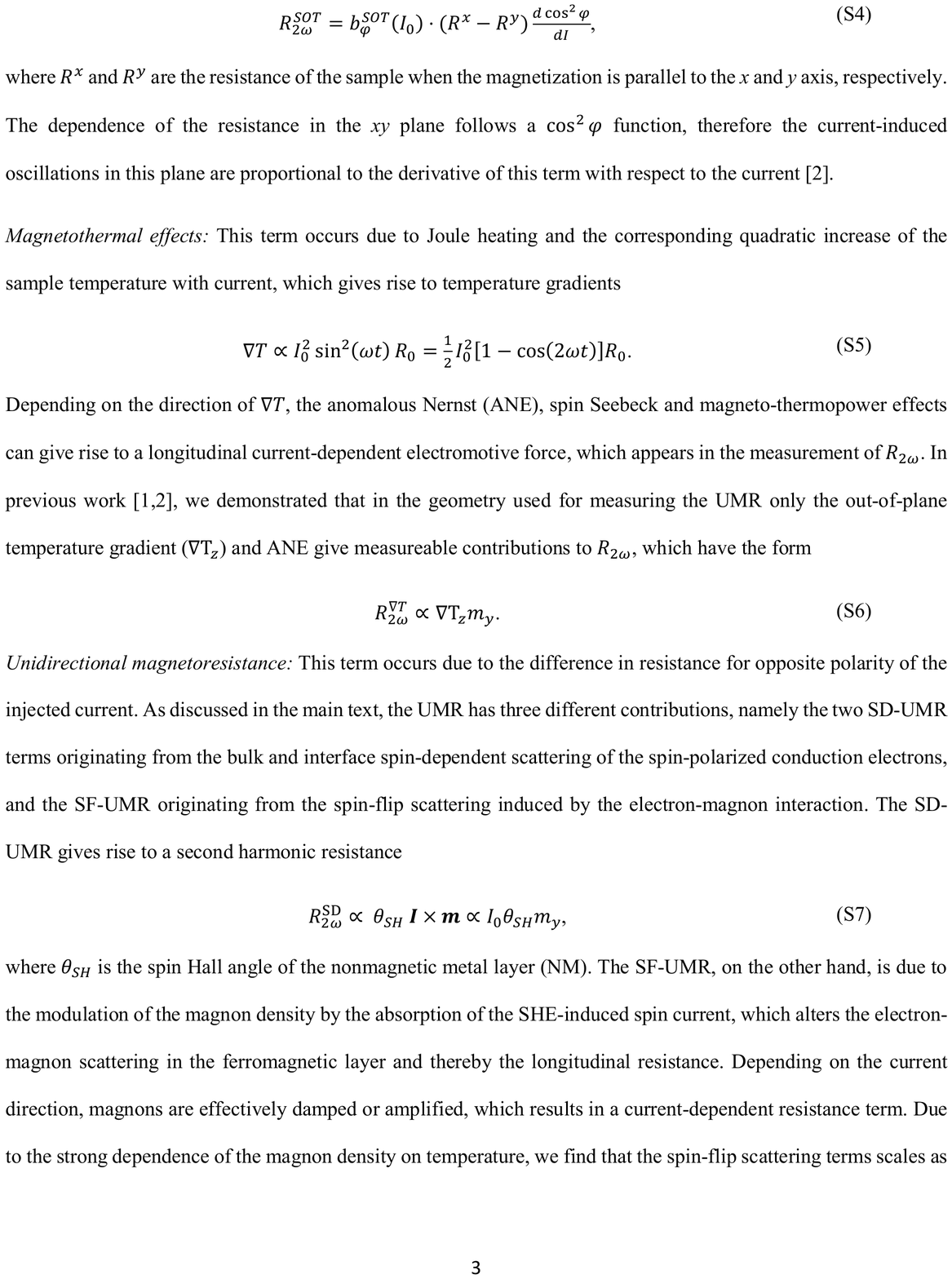}
\includepdf[pages=-]{Avci_SM_Part3.pdf}
\includepdf[pages=-]{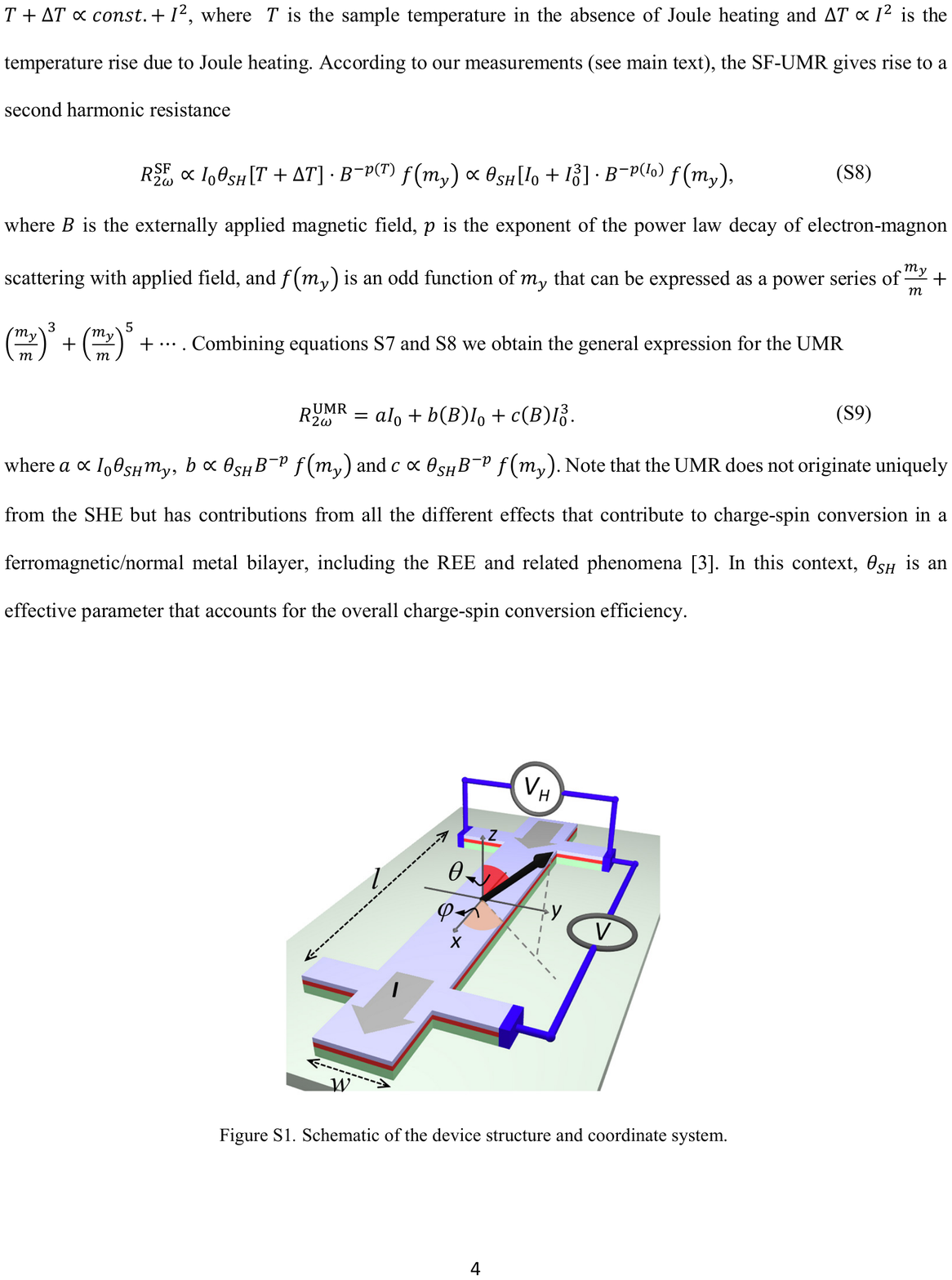}
\includepdf[pages=-]{Avci_SM_Part4.pdf}
\includepdf[pages=-]{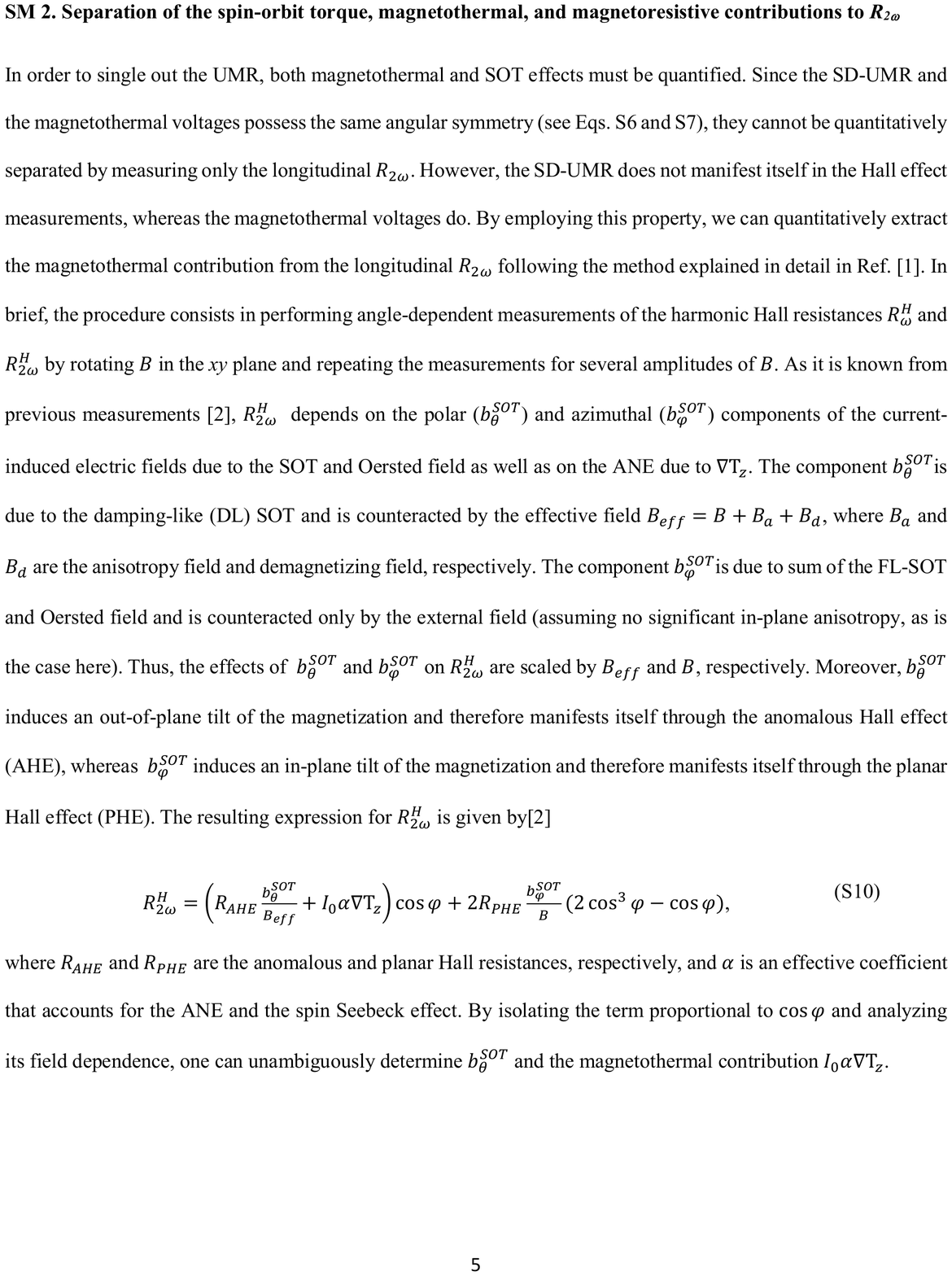}
\includepdf[pages=-]{Avci_SM_Part5.pdf}
\includepdf[pages=-]{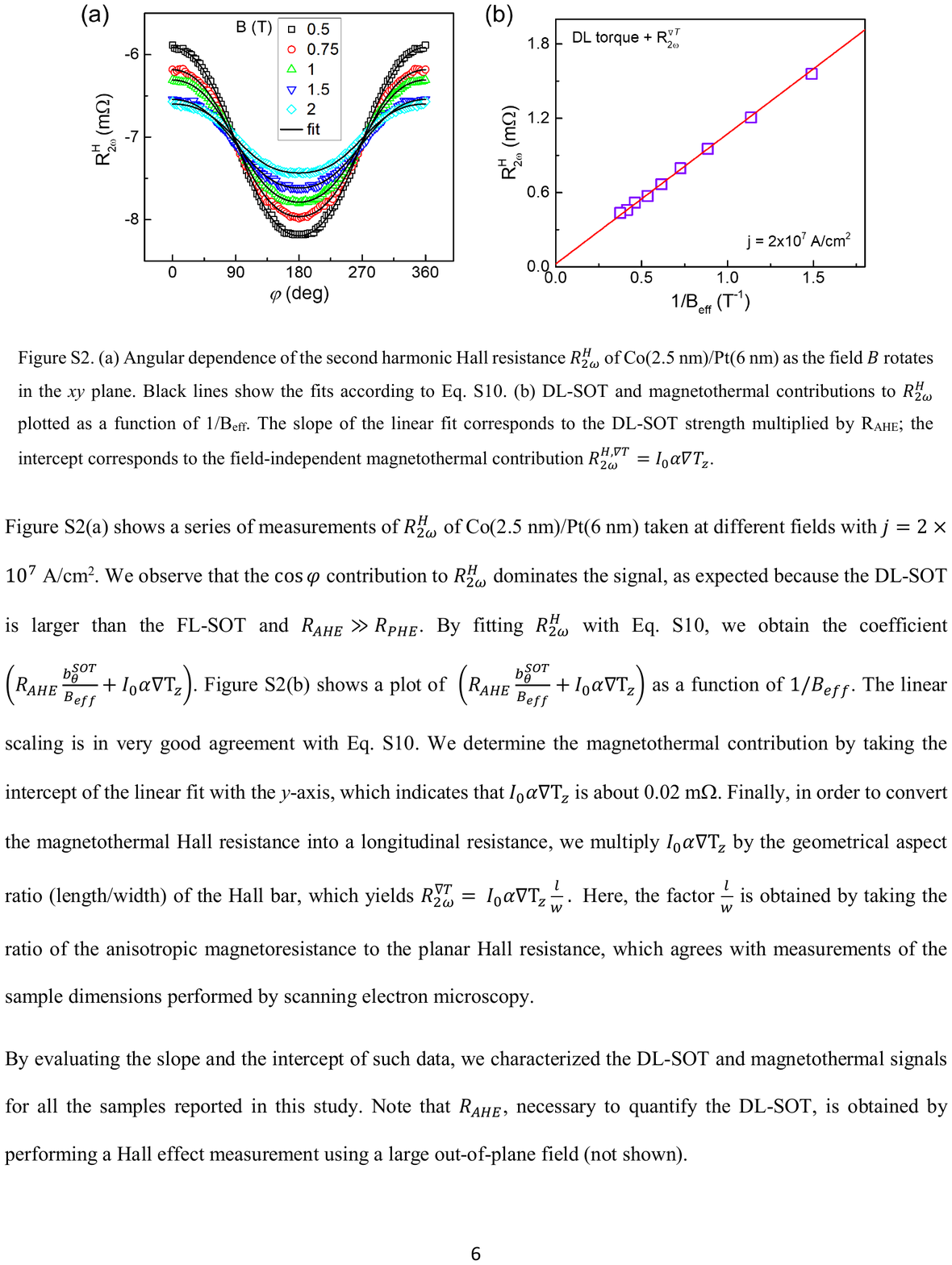}
\includepdf[pages=-]{Avci_SM_Part6.pdf}
\includepdf[pages=-]{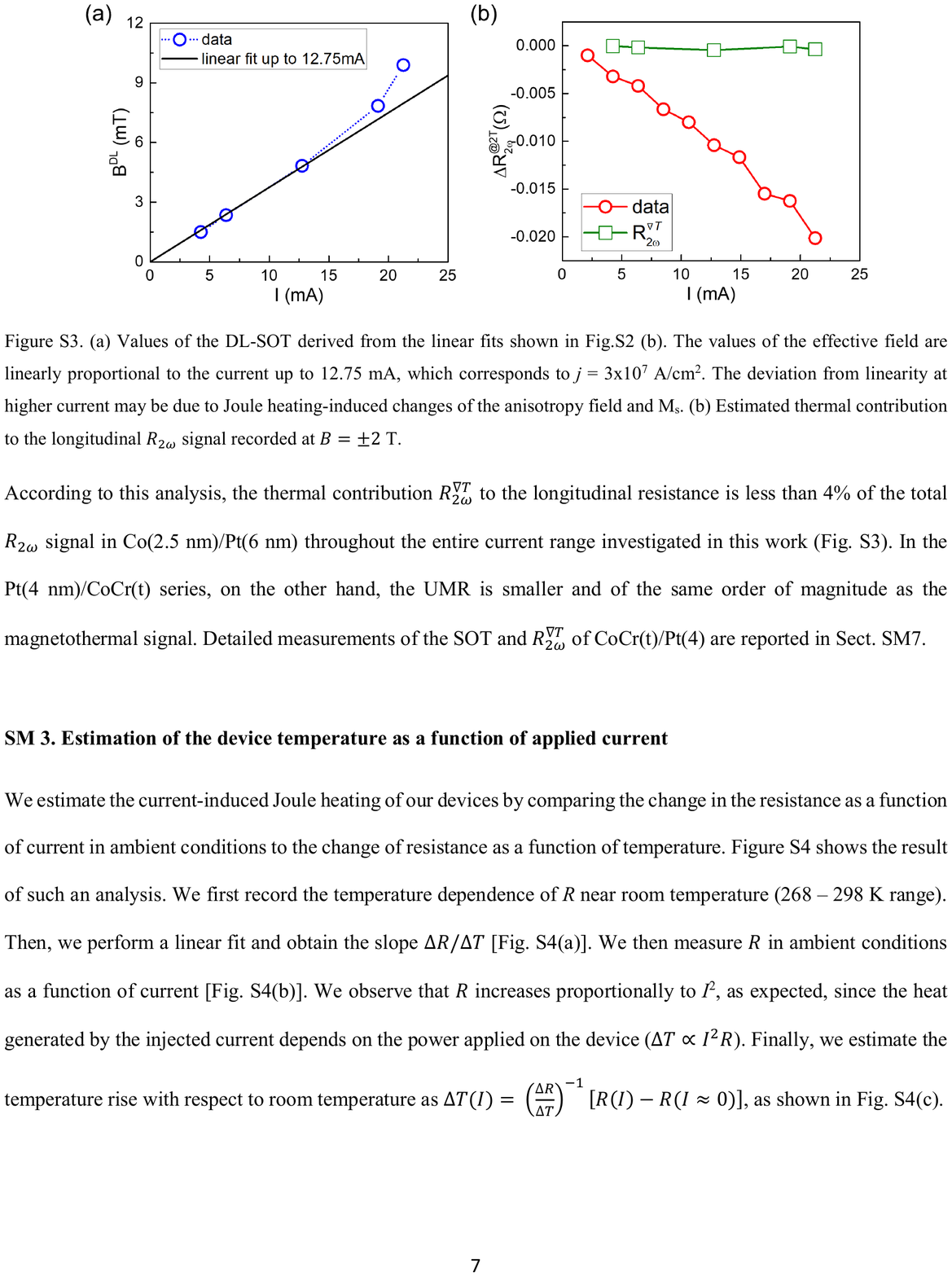}
\includepdf[pages=-]{Avci_SM_Part7.pdf}
\includepdf[pages=-]{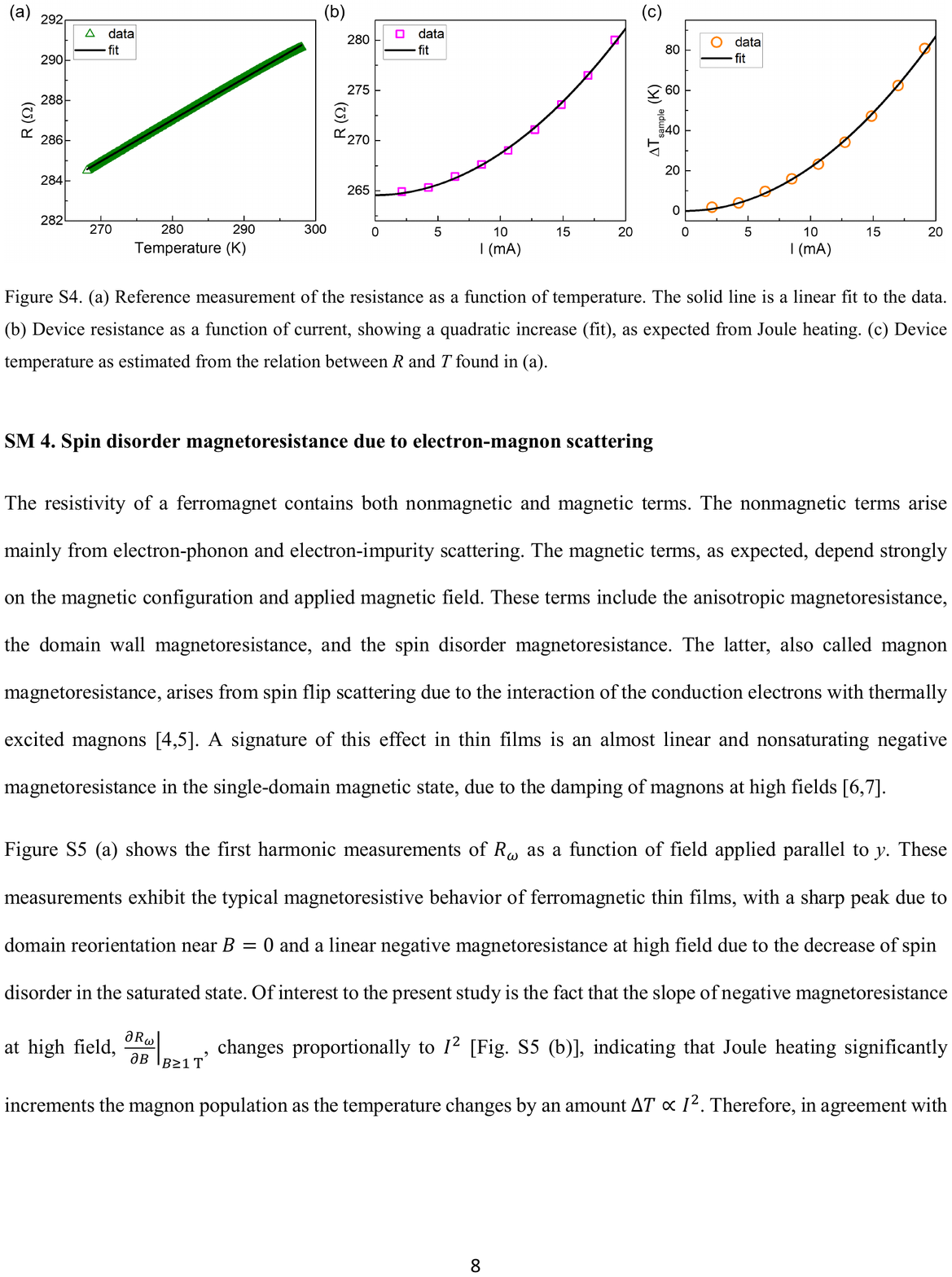}
\includepdf[pages=-]{Avci_SM_Part8.pdf}
\includepdf[pages=-]{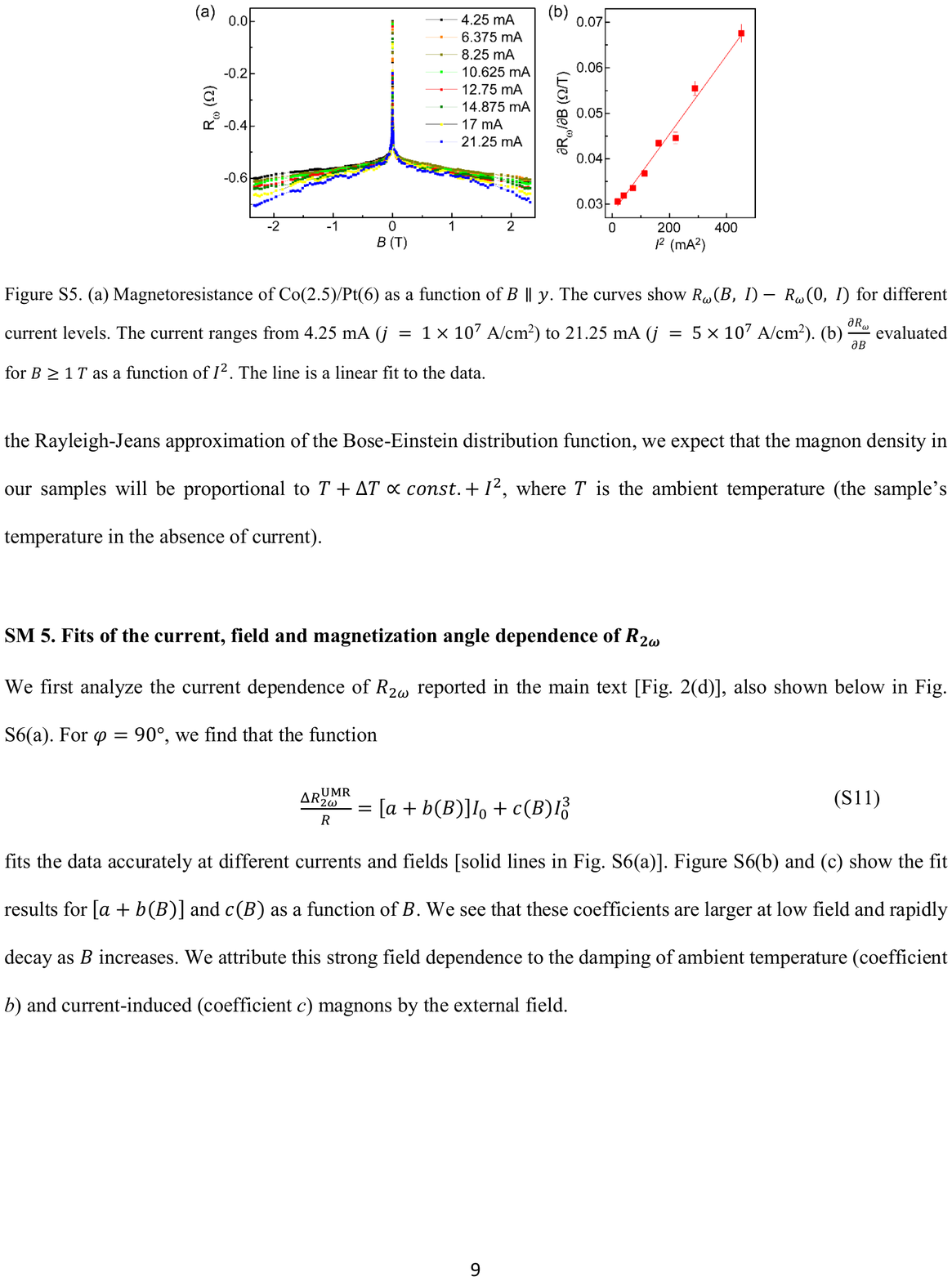}
\includepdf[pages=-]{Avci_SM_Part9.pdf}
\includepdf[pages=-]{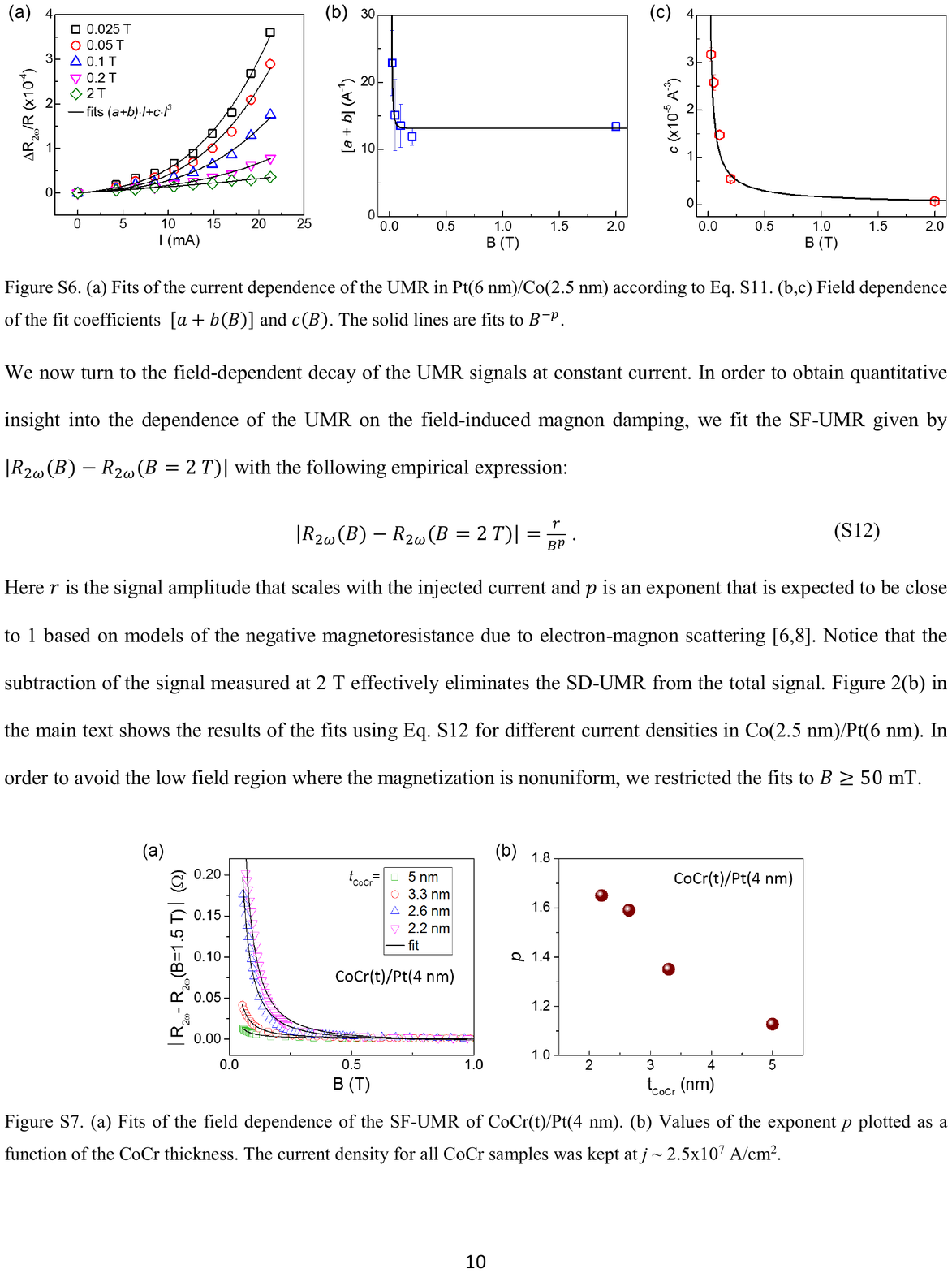}
\includepdf[pages=-]{Avci_SM_Part10.pdf}
\includepdf[pages=-]{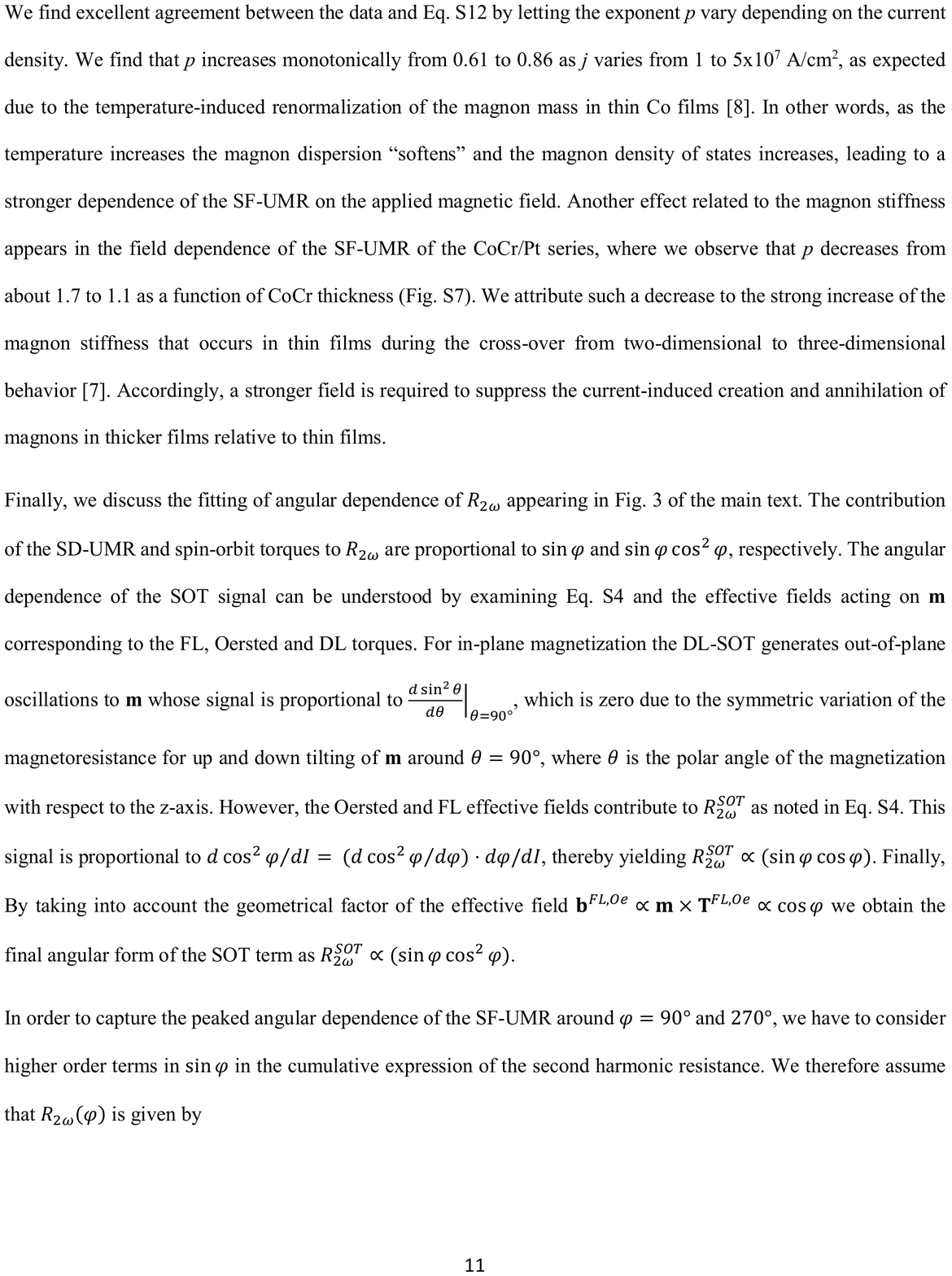}
\includepdf[pages=-]{Avci_SM_Part11.pdf}
\includepdf[pages=-]{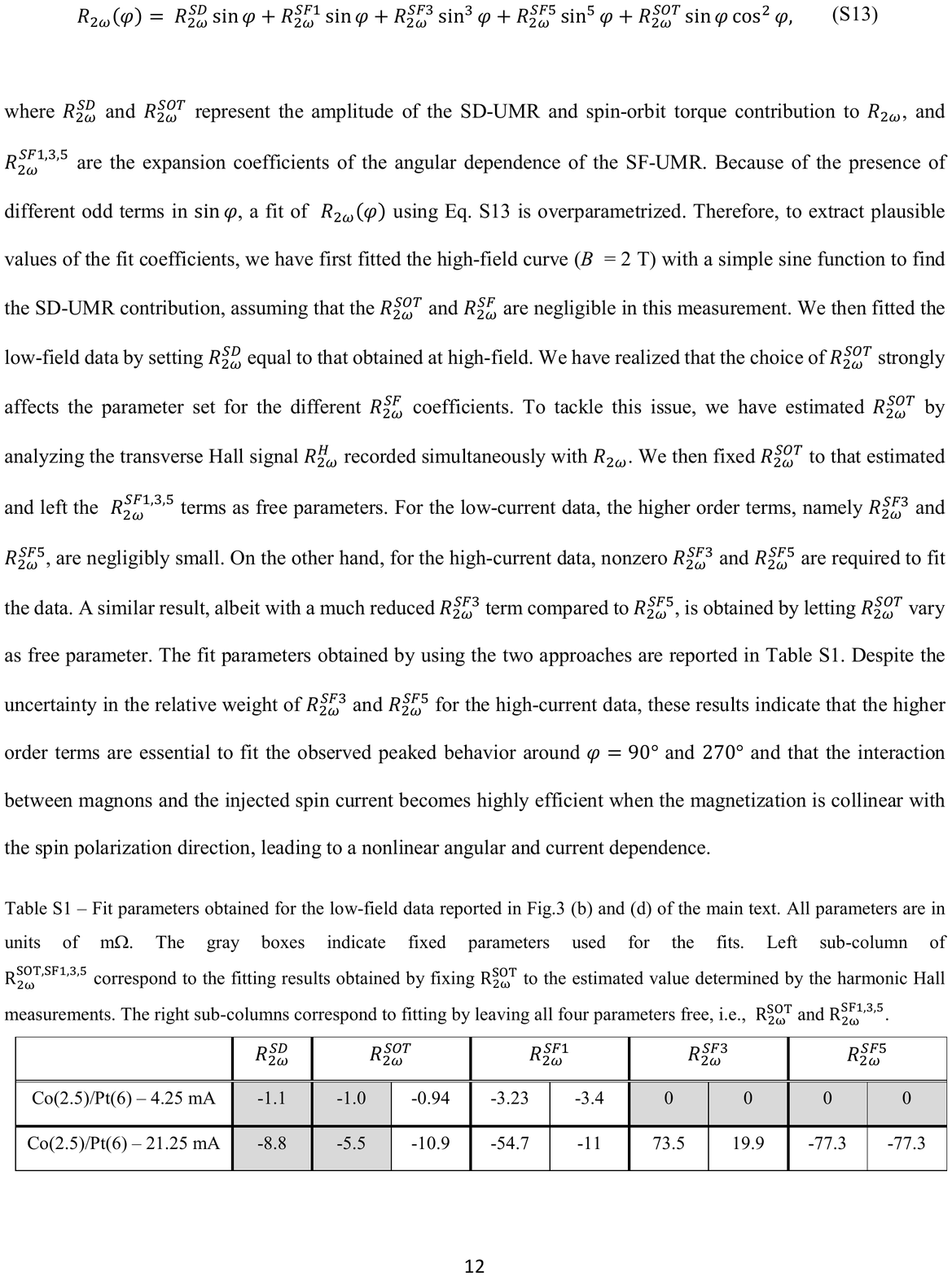}
\includepdf[pages=-]{Avci_SM_Part12.pdf}
\includepdf[pages=-]{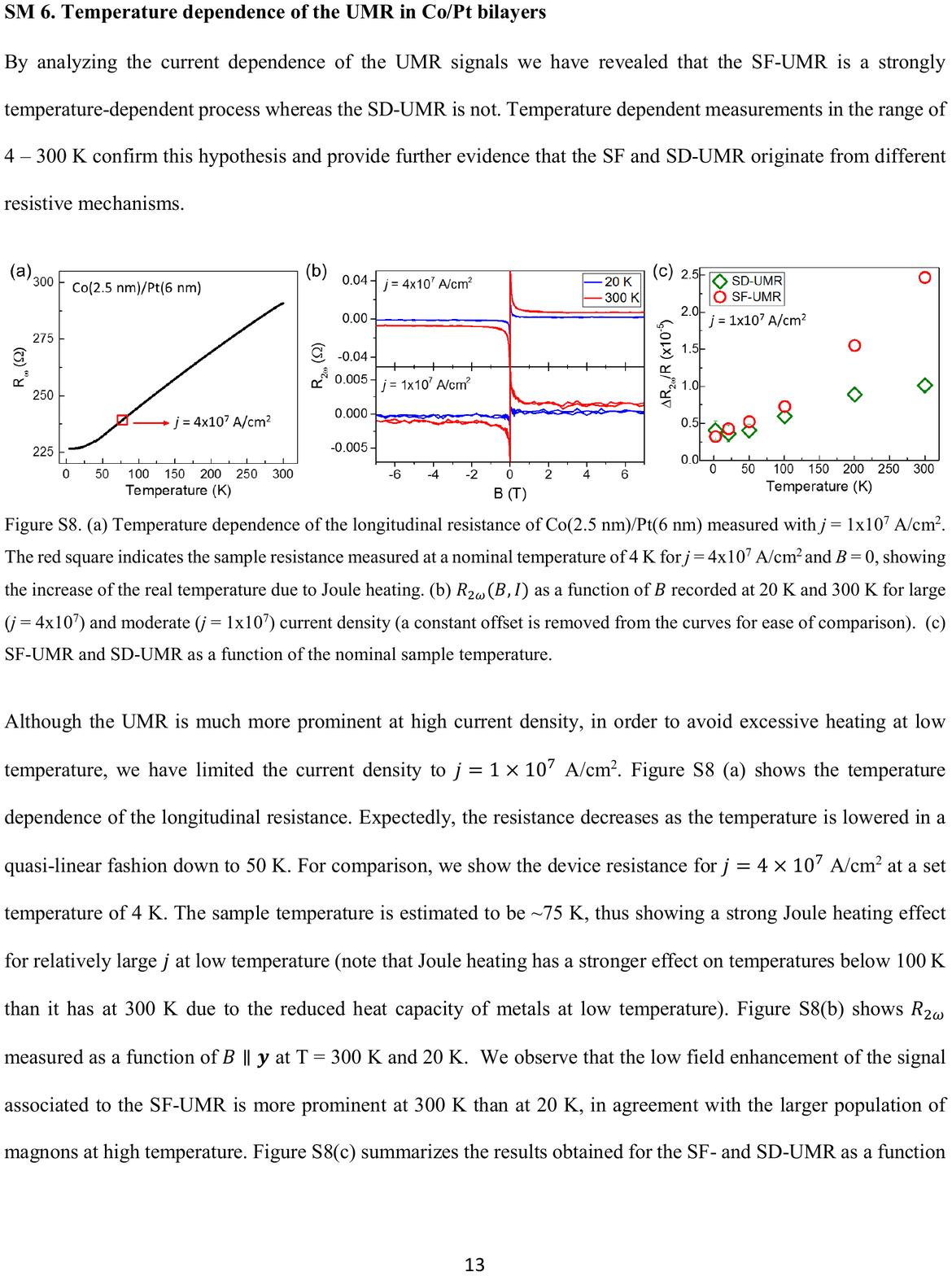}
\includepdf[pages=-]{Avci_SM_Part13.pdf}
\includepdf[pages=-]{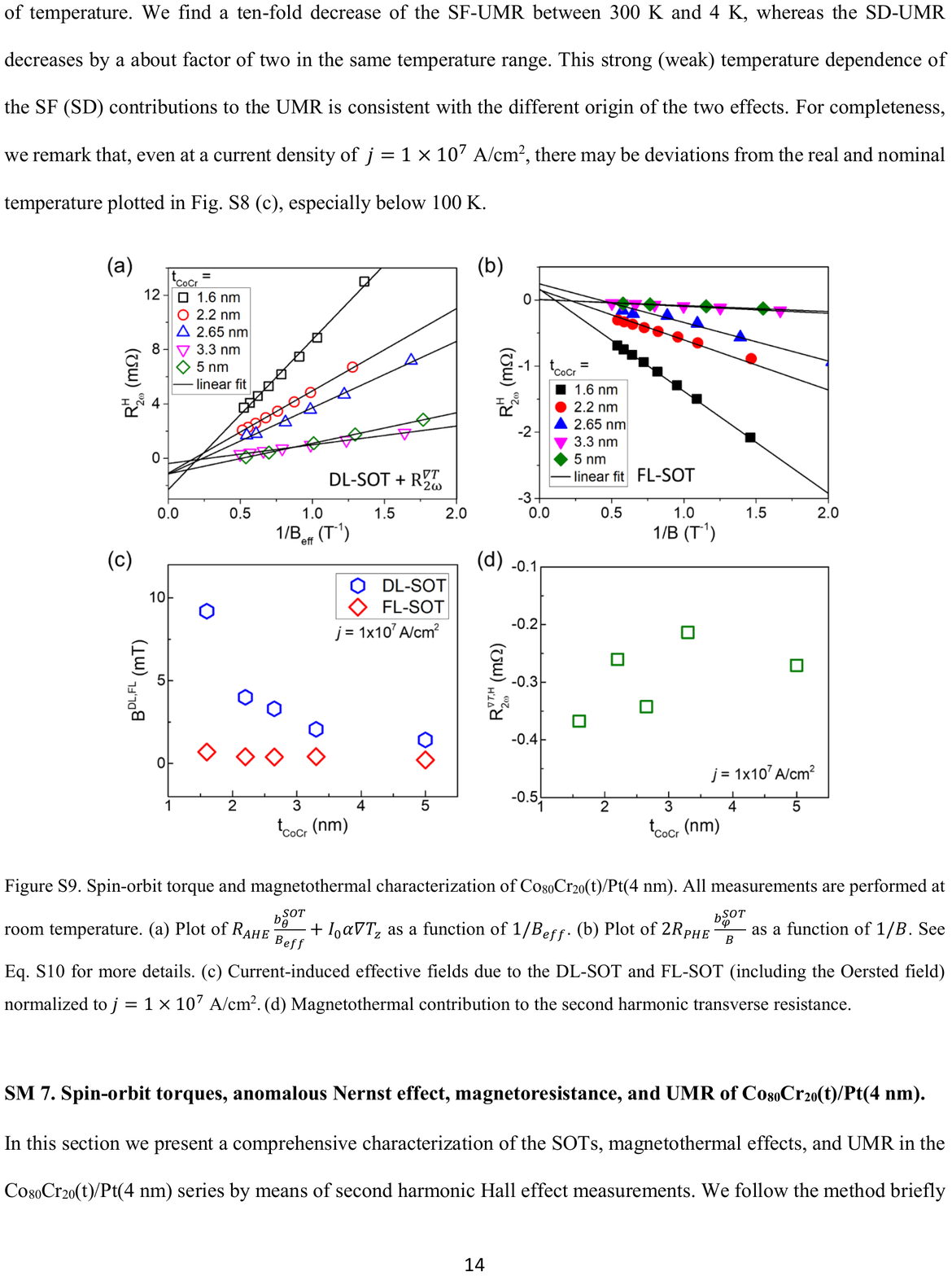}
\includepdf[pages=-]{Avci_SM_Part14.pdf}
\includepdf[pages=-]{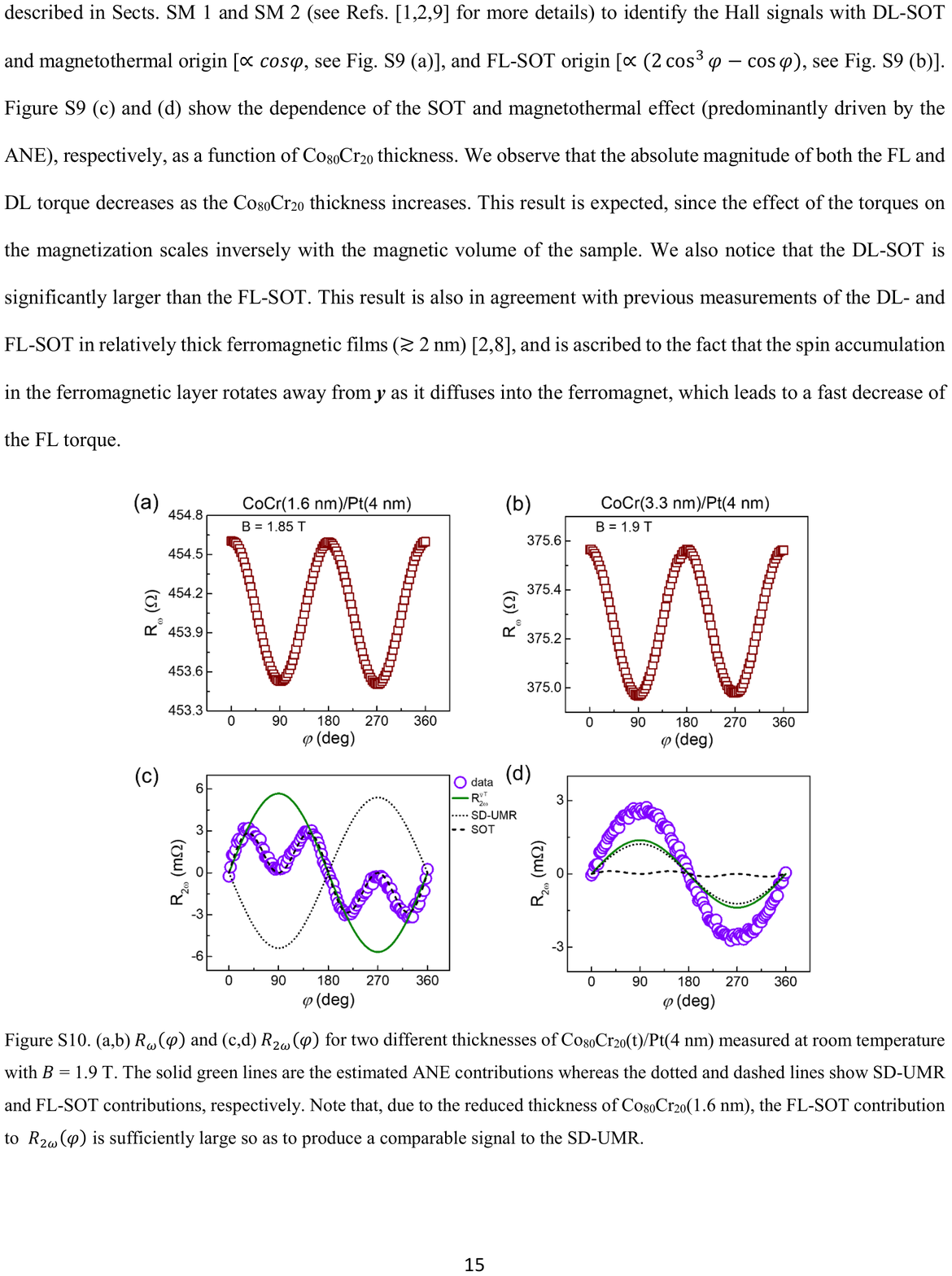}
\includepdf[pages=-]{Avci_SM_Part15.pdf}
\includepdf[pages=-]{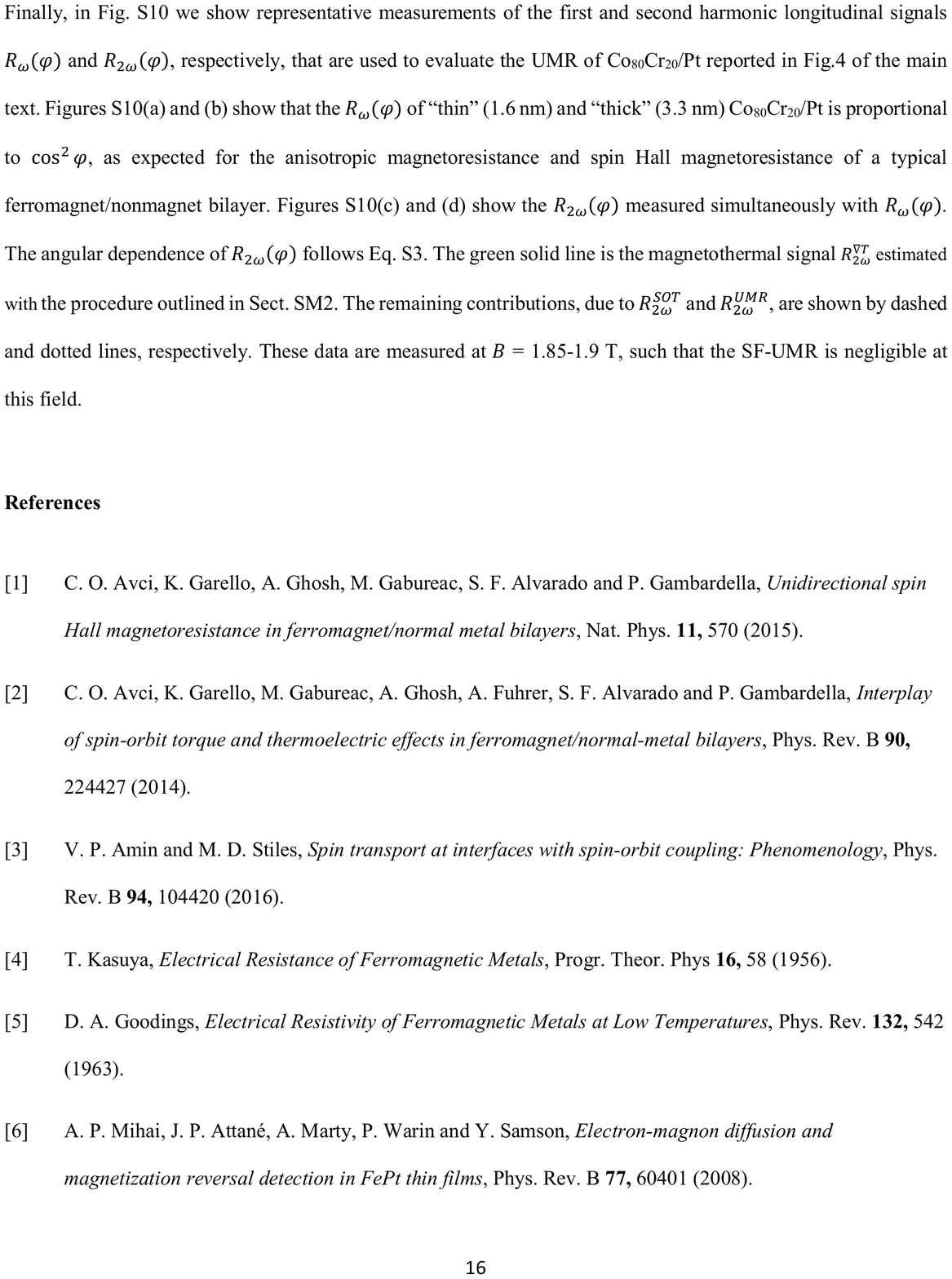}
\includepdf[pages=-]{Avci_SM_Part16.pdf}
\includepdf[pages=-]{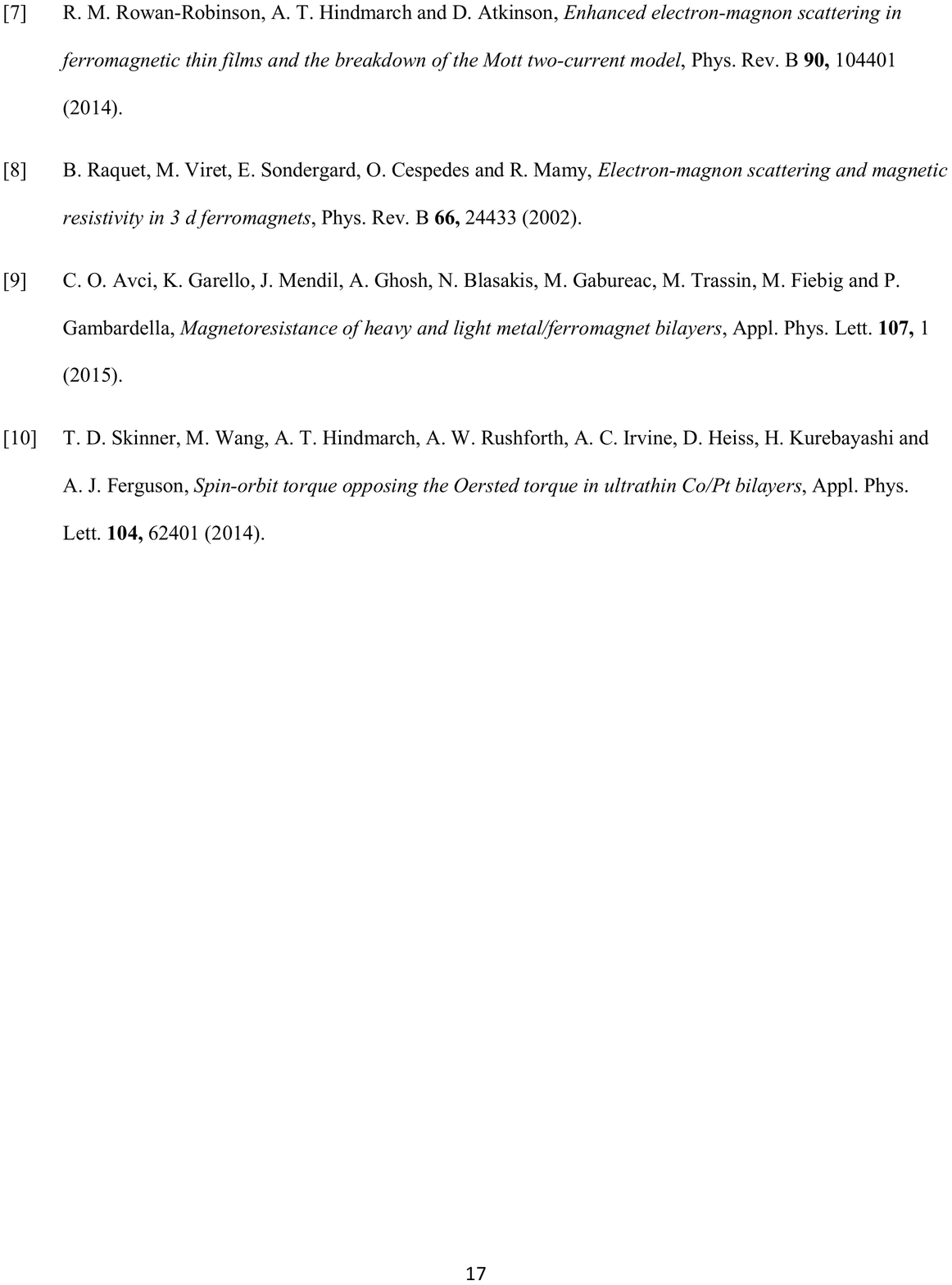}
\includepdf[pages=-]{Avci_SM_Part17.pdf}